\documentclass[pre,preprint]{revtex4}

\usepackage{amsfonts}
\usepackage{amsmath}

\usepackage{theorem}

\newcommand{\one}{\mbox{\tt 1}\hspace{-0.057 in}\mbox{\tt l}}
\newcommand{\natuerl}{{\mathbb N}}

\newcommand{\Ket}[1]{|#1\rangle}

\newcommand{\Tr}{\mbox{\rm\small Tr\ }}

\newcommand{\balphas}{\mbox{\boldmath $\scriptstyle \alpha$}}

\newcommand{\bbetas}{\mbox{\boldmath $\scriptstyle \beta$}}

\theoremstyle{break}

\begin{document}

\title{A simple necessary decoherence condition for a set  
of histories}
\author{Artur Scherer}
\author{Andrei N.\ Soklakov\footnote{the corresponding author:
a.soklakov@rhul.ac.uk; Tel: +44(0)1784 44 3106; Fax: +44(0)1784 430766}}
\author{R\"udiger Schack}
\affiliation{Department of Mathematics, Royal Holloway, University of London,
Egham, Surrey, TW20 0EX, UK.}
\date{20 January 2004}
\begin{abstract}
Within the decoherent histories formulation of quantum mechanics, we
investigate necessary conditions for decoherence of arbitrarily long
histories. We prove that fine-grained histories of arbitrary
length decohere for all classical initial states if and only if the unitary
evolution preserves classicality of states (using a natural formal definition
of classicality). We give a counterexample showing that this equivalence
does not hold for coarse-grained histories.\\


\end{abstract}
\maketitle

The formalism of decoherent histories was introduced to provide a
self-contained description of closed quantum systems
\cite{Griffiths1984,Omnes1988,Gell-Mann1990,Dowker1992,Gell-Mann1992}.
Applications include, e.g., quantum cosmology \cite{Hartle1997}, a
derivation of the equations of classical 
hydrodynamics \cite{Halliwell_1998},
and
the coarse-grained evolution of iterated quantum maps \cite{Soklakov2002}. The
concept of {\em histories\/} is central to this approach. A history is defined
to be a time-ordered sequence of quantum mechanical ``propositions''.
Due to quantum interference, one cannot always assign
probabilities to a set of histories in a consistent way. For this to be
possible, the set of histories must be decoherent.

In general it is very difficult to decide if a given set of histories is
decoherent.  As the length of the histories increases, checking the
decoherence conditions soon becomes extremely cumbersome. This is
especially true when the system dynamics is difficult to simulate as, e.g., in
the case of a chaotic quantum map.  In this paper we
investigate a simple criterion for decoherence.  Since this
criterion can be shown trivially to be a sufficient condition for decoherence,
the analysis below concentrates on the question of whether the criterion is a
necessary condition as well.

The paper is organized as follows. We begin with defining our setting within
the framework of the decoherent histories formulation of quantum mechanics.
We state our main results in the form of two theorems. We then prove the
theorems and conclude with a short discussion.

\noindent{\bf Definition 1:}
A set of projectors $\{P_{\mu}\}$ on a Hilbert space $\cal H$ is 
called a projective {\em partition} of $\cal H$, if $\:\forall\, \mu, 
\mu'\,:\;\:P_{\mu}P_{\mu'}=\delta_{\mu\mu'}P_{\mu}\:$ and $\:\sum_{\mu}P_{\mu}
=\one_{\cal H}$. Here, $\one_{\cal H}$ denotes the unit operator. 
We will call a projective partition  {\em
  fine-grained\/} if all projectors are one-dimensional, 
i.e.,  $\,\forall\,\mu\;$
$\mbox{dim}\big(\mbox{supp}(P_{\mu})\big)=1$
\footnote{The {\em support} of a Hermitian operator $A$ is defined to 
be the vector space spanned by the eigenvectors of $A$ corresponding 
to its non-zero eigenvalues.},
and {\em coarse-grained} otherwise.
\vspace{2mm}

\noindent{\bf Definition 2:}
A state represented by the density operator $\rho$ is called 
{\em classical with respect to (w.r.t.) a partition  $\{P_{\mu}\}$}
of the Hilbert space $\cal H$, if 
\begin{equation}
\rho=\sum_k p_k \rho_k\;,\quad\mbox{where}\quad \forall k\, \exists \mu
 \quad\mbox{such that}\quad \Tr[P_{\mu}\rho_k]=1\;.
\end{equation}

The last statement means that for every $\rho_k$ in the decomposition 
$\rho=\sum_k p_k \rho_k$ there exists a $P_{\mu}\in\{P_{\nu}\}$ such 
that $\mbox{supp}(\rho_k)\subseteq\mbox{supp}(P_{\mu})$.
We denote by $\mathcal{S}$ the set of all 
density operators on $\cal H$, and by 
$\mathcal{S}^{\mbox{\scriptsize cl}}_{\{P_{\mu}\}}$ the set 
of all density operators that are classical w.r.t.\ $\{P_{\mu}\}$.
\vspace{2mm}

\noindent{\bf Definition 3:}
Given a projective partition $\{P_{\mu}\}$ of a Hilbert space $\cal H$, 
we denote by $\mathcal{K}[\{P_{\mu}\}\,;\,k\,]:=\big\{h_{\balphas}\,:\:
h_{\balphas}=\left(P_{\alpha_{t_1}},P_{\alpha_{t_2}},
 \dots, P_{\alpha_{t_k}}\right)\in\{P_{\mu}\}^k\big\}$ the 
corresponding exhaustive set of mutually exclusive histories of length
$k$. Histories are thus defined to be time-ordered sequences of projection
operators, corresponding to quantum-mechanical propositions. Here we 
assume that the projectors $P_{\alpha_{t_j}}$ are chosen
from the same partition for all times $t_j$,  $j=1,\ldots,k$.

\vspace{2mm}

An initial state $\rho\in\mathcal{S}$ and a unitary  
dynamics generated by a unitary map $U:\cal H\rightarrow\cal H$ induce 
a probabilistic structure on the event algebra 
associated with $\mathcal{K}[\{P_{\mu}\}\,;\,k]$, 
if certain consistency conditions are fulfilled.
These are given in terms of properties of the {\em decoherence functional\/}
$\mathcal{D}_{U,\,\rho}\,[\cdot,\cdot]$ on 
$\mathcal{K}[\{P_{\mu}\}\,;\,k\,]\times
\mathcal{K}[\{P_{\mu}\}\,;\,k\,]\,$, defined by 
\begin{equation} 
\mathcal{D}_{U,\,\rho}\,[h_{\balphas},h_{\bbetas}]:=
\mbox{Tr}\left[C_{\balphas}\,\rho\,C_{\bbetas}^{\dagger}\right]\:,
\end{equation}
where 
\begin{equation} 
C_{\balphas} := 
U^{\dagger\,k}P_{\alpha_k}UP_{\alpha_{k-1}}U\dots
  P_{\alpha_2}UP_{\alpha_1}U\;.
\end{equation}
The set $\mathcal{K}[\{P_{\mu}\}\,;\,k\,]$ is said to be 
{\em decoherent\/} or {\em consistent\/} with respect to a given 
unitary map $U:\cal H\rightarrow\cal H$ and a given initial 
state $\rho\in\mathcal{S}$, if 
\begin{equation}  \label{eq:consistency}
\mathcal{D}_{U,\,\rho}\,[h_{\balphas},h_{\bbetas}]
\propto \delta_{\balphas\bbetas}\equiv
\prod_{j=1}^k\delta_{\alpha_j \beta_j}
\end{equation}
for all $h_{\balphas},h_{\bbetas}\in\mathcal{K}[\{P_{\mu}\}\,;\,k\,]$.
These are the consistency conditions. 
If they are fulfilled, probabilities  
may be assigned to the histories   
and are given by the diagonal elements 
of the decoherence functional, 
$p[h_{\balphas}]=\mathcal{D}_{U,\,\rho}\,[h_{\balphas},h_{\balphas}]$.

What we have just described is a slightly simplified version of the 
general decoherent histories formalism. In general, both the partition and the
unitary may depend on time. Furthermore, 
several consistency conditions of different strength are considered in 
the literature~\cite{Gell-Mann_1998}. The conditions given above
are known as {\em medium decoherence\/} \cite{Gell-Mann1992}. 

The mathematical framework used here is similar to the formalism of symbolic
dynamics \cite{Alekseev1981,Badii1997}.  As in the theory of classical
dynamical systems we start by partitioning the space of possible system
states, using a fixed partition for all times. We proceed by looking for a
probability measure over the set of histories---again in close analogy with
symbolic dynamics. This analogy has been exploited before in a symbolic
dynamics approach to the quantum baker's map
\cite{Soklakov2002,Soklakov2000a}.

Although, at a fundamental level, the decoherent histories 
approach does not need the notion of a measurement, 
this notion can be very helpful for visualizing the 
properties of quantum states. For example, 
a projective partition can be regarded as defining a 
projective measurement on the system. 
One can see that classical states are 
not perturbed by such measurements. 
Indeed, one can easily show that a state $\rho\in\mathcal{S}$ is classical 
with respect to the partition $\{P_{\mu}\}$  if, and only if,  
$\;\sum_{\mu}P_{\mu}\,\rho\, P_{\mu}=\rho\:$.  
This property motivates the name ``classical states''. 
In the theorems stated below we will always choose classical states
as the initial states for the histories. This choice is motivated by the fact
that only classical states 
$\rho\in\mathcal{S}^{\mbox{\scriptsize cl}}_{\{P_{\mu}\}}$ 
can be \lq\lq prepared'' by the projective measurement
 defined by  $\{P_{\mu}\}$. 

In this paper we make progress towards finding a simple 
characterization of the set of unitaries that, given a classical
initial state, lead to decoherent histories of arbitrary length.
Imagine a unitary evolution that transforms every classical 
state into a classical state. If the initial state is classical, 
this evolution trivially leads to decoherent histories.
One can easily see that in this case the decoherence functional
is diagonal for histories of any length. 
It is not immediately clear, however, whether any unitary that
leads to the desired decoherence effect must preserve
classicality of states. In what follows we show that this is the case
only for fine-grained histories.

\noindent{\bf Theorem 1:}
{\it Let a fine-grained projective partition}  $\{P_{\mu}\}$ 
{\it of a finite dimensional Hilbert space $\cal H$ and a unitary 
map} $U$ {\it on} $\cal H$ {\it be given. The decoherence conditions are
then satisfied for all classical initial states and arbitrarily long
histories if and only if $U$ preserves classicality of states, i.e.,}
\begin{equation}  \label{eq:decoherence}
\forall \,\rho\in\mathcal{S}^{\mbox{\scriptsize cl}}_{\{P_{\mu}\}}\;
\forall\, k\in\hspace{-0.5mm}\natuerl \;\;\forall\,h_{\balphas},
h_{\bbetas}\in\mathcal{K}[\{P_{\mu}\}\,;\,k\,]:
\;\mathcal{D}_{U,\rho}\,[h_{\balphas},h_{\bbetas}]
\propto  \delta_{\balphas\bbetas}
\end{equation}
{\em if and only if\/}
\begin{equation}  \label{eq:classicality}
\forall
  \,\rho\in\mathcal{S}^{\mbox{\scriptsize cl}}_{\{P_{\mu}\}}
\,:\;\;U\rho U^{\dagger}\,\in 
\mathcal{S}^{\mbox{\scriptsize
    cl}}_{\{P_{\mu}\}}  \;.
\end{equation}

\vspace{3mm}

\noindent{\bf Theorem 2:}
{\it For coarse-grained partitions, the classicality 
condition~(\ref{eq:classicality}) of Theorem 1 is 
in general not a necessary condition. More precisely, 
there exists a coarse-grained projective partition and 
a unitary map such that the classicality 
condition~(\ref{eq:classicality}) is not 
satisfied but the decoherence
condition~(\ref{eq:decoherence}) is valid}. 

\vspace{3mm}

Thus, decoherence for arbitrarily long histories and classical initial states
is a sufficient condition for $U$ to preserve classicality of states in the
fine-grained case, but not in the coarse-grained case. In general, decoherence
does not imply that the unitary evolution preserves classicality.

In our theorems, the decoherence condition is formulated for 
any $k\in\natuerl$, i.e., 
arbitrary history lengths, corresponding to 
an arbitrary number of iteration steps of the unitary 
map $U$. This is a very strong condition.
It can be relaxed if the Hilbert space is two-dimensional. In this case,
decoherence of all histories of length $k=2$ 
for all classical initial states is equivalent to the condition
that the unitary evolution preserves classicality of states.

In general, however, it is not sufficient to restrict attention to histories
  of a fixed finite length. This is made precise in the following example. 
For a given $K\in\natuerl$ consider 
a Hilbert space $\cal H$ with dimension $d=2K$. Let  
$\{P_{\mu}=|\mu\rangle\langle\mu|\;:\;\mu=0,1,\dots, 
d-1\,\}$ be a fine-grained partition of $\cal H$, where 
the kets $|\mu\rangle$ form an orthonormal basis of 
$\cal H$. Define a unitary 
map $U:\cal H\rightarrow\cal H$ by 
\begin{eqnarray}
|0\rangle&\rightarrow &U|0\rangle
=\frac{1}{\sqrt{2}}\,(|2\rangle+|3\rangle)\cr
|1\rangle&\rightarrow &U|1\rangle
=\frac{1}{\sqrt{2}}\,(|2\rangle-|3\rangle)\cr
|\nu\rangle&\rightarrow
&U|\nu\rangle=|\nu+2\rangle\quad\mbox{for}\quad \nu=2,3,\dots,(d-3)\cr
|d-2\rangle&\rightarrow &U|d-2\rangle=|0\rangle\cr
|d-1\rangle&\rightarrow &U|d-1\rangle=|1\rangle\;.
\end{eqnarray}
The map $U$ does not preserve classicality w.r.t.\ $\{P_{\mu}\}$.  For $k>K$
and, e.g., the classical initial state $\rho=|0\rangle\langle0|
\in\mathcal{S}^{\mbox{\scriptsize cl}}_{\{P_{\mu}\}}$, the set of histories
$\mathcal{K}[\{P_{\mu}\}\,;\,k\,]$ does not decohere.  One can easily show,
however, that $\mathcal{K}[\{P_{\mu}\}\,;\,k\,]$ decoheres for all
$\rho\in\mathcal{S}^{\mbox{\scriptsize cl}}_{\{P_{\mu}\}}$ and all $k\le K$.
We have thus found, for any $K\in\natuerl$, an example in
which $U$ does not preserve classicality, but the decoherence condition is
satisfied for all classical initial states and all histories up to length $K$.

\noindent
In the proof of theorem 1, we will use the following lemma:

\noindent 
{\bf Lemma:}
Let $\mathcal{H}$ be a {\em finite dimensional} Hilbert space, and let
$U$ be a unitary map on ${\cal H}$.
Then $\forall\,\epsilon >0\;\;\exists\,
q\in\natuerl$ such that $\parallel\! U^q-\one_{\cal H} \!\parallel 
< \epsilon\,$, where $\parallel\! \cdot \!\parallel$ denotes the 
conventional operator norm, $\parallel\hspace{-1mm} A\hspace{-1mm} \parallel
=\sup\{\parallel\! Av \!\parallel\::\: 
v\in\mathcal{H}\,,\,\parallel\! v\!\parallel=1\}$ 
for any operator $A$ on $\mathcal{H}$.

\noindent
{\bf Proof of the Lemma:}
Since our Hilbert space is finite dimensional, $U$ has
a discrete eigenvalue spectrum. All eigenvalues 
of a unitary operator have modulus 1. 
The spectral decomposition of $U$ can therefore 
be written in the form
\begin{equation}
U=\sum_{j=1}^d  e^{2\pi i \,\xi_j}\,|\Omega_j\rangle\langle\Omega_j|\;,
\end{equation}
where $d:=\mbox{dim}(\mathcal{H})$, $\xi_1,\dots, \xi_d \,$ are   
real numbers, and $|\Omega_j\rangle$ are the eigenvectors of $U$. 
The Lemma is trivially true if $\xi_1,\dots, \xi_d\,$ are all rational. In 
this case we immediately get $U^{q}=\one_{\cal H}$, if $q$ is a common 
denominator of $\xi_1,\dots, \xi_d \,$.  For arbitrary
$\xi_1,\dots, \xi_d$, we make use of 
a number-theoretical result, known as {\em Dirichlet's theorem on 
simultaneous diophantine approximation\/} \cite{Cassels1957}.  
We wish to get a simultaneous approximation of $\xi_1,\dots, \xi_d \,$ by 
fractions
\begin{equation}
\frac{p_1}{q},\frac{p_2}{q},\dots, 
\frac{p_d}{q}
\end{equation}
with a common denominator $q$. 
Furthermore we wish to have the ability to choose 
the common denominator $q$ in such a way that 
$\mbox{max}\{|q\xi_1-p_1|,\dots, |q\xi_d-p_d|\}$
becomes arbitrarily small. According to Dirichlet's 
theorem this is possible: {\em If $\xi_1,\dots, \xi_d \,$ 
are any real numbers such that at least one of them 
is irrational, then the system of inequalities} 
\begin{equation}
\left|\xi_j- \frac{p_j}{q}\right|<\frac{1}{q^{1+\frac{1}{d}}}\quad
\mbox{\em with}
\quad q,p_j\in\natuerl \quad (j=1,2,\dots,d)
\end{equation}
{\em has infinitely many solutions. In particular,}  
$\mbox{max}\{|q\xi_1-p_1|,\dots, |q\xi_d-p_d|\}<
q^{-\frac{1}{d}}$ {\em holds for infinitely many 
integers} $q\in\natuerl$. As a consequence, given any 
$\epsilon>0$, we can always find an integer $q\in\natuerl$ 
so that, for every $j\in\{1,2,\dots,d\}$, the product $q\xi_j$ differs from 
an integer by less than $ \epsilon$. 

To prove the Lemma, let any $\epsilon>0$ be given. 
Define $\epsilon':=\frac{\epsilon}{d(e^{2\pi}-1)}$. 
According to 
Dirichlet's Theorem there always exists a  
$q=q(\epsilon')\in\natuerl$ such that, for every $j$,  
$q\xi_j$ differs from an integer
by less than $ \epsilon'$. It follows that 
\begin{equation}
U^q = \sum_{j=1}^d  e^{2\pi i \,q\,\xi_j}\,|\Omega_j\rangle\langle\Omega_j|
=\sum_{j=1}^d  e^{2\pi i
   \,\epsilon_j}\,|\Omega_j\rangle\langle\Omega_j|
\end{equation}
with some very small numbers $\epsilon_j$ satisfying $|\epsilon_j|< \epsilon'$ 
for all $j$. Hence
\begin{eqnarray}
\parallel  U^q-\one_{\cal H} \parallel &=&
\parallel\sum_{j=1}^d (e^{2\pi i
\epsilon_j}-1)|\Omega_j\rangle\langle\Omega_j|\parallel\nonumber\\
&\le&\sum_{j=1}^d
\sum_{\nu=1}^{\infty}\frac{(2\pi)^{\nu}}{\nu!}\left|\epsilon_j\right|^{\nu}
\underbrace{\parallel\,|\Omega_j\rangle\langle\Omega_j|\,\parallel}_{=1}
\nonumber\\
&<&\sum_{j=1}^d \sum_{\nu=1}^{\infty}\frac{(2\pi)^{\nu}}{\nu!}\epsilon'^{\hspace{0.1mm}\nu}
<\sum_{j=1}^d \sum_{\nu=1}^{\infty}\frac{(2\pi)^{\nu}}{\nu!}\epsilon'
\nonumber\\
&=& d\cdot\epsilon'\cdot(e^{2\pi}-1)=\epsilon \;.
\end{eqnarray}
This proves the Lemma.

\noindent{\bf Proof of Theorem 1:}\\
The classicality condition~(\ref{eq:classicality}) implies the decoherence
condition~(\ref{eq:decoherence}) trivially. We will prove the converse by
contradiction, i.e., we will assume that the 
classicality condition~(\ref{eq:classicality}) is not satisfied, 
and then show that this assumption contradicts the decoherence 
condition~(\ref{eq:decoherence}).

Assume condition~(\ref{eq:classicality}) is not satisfied. 
This means there exists a classical state 
$\rho\in\mathcal{S}^{\mbox{\scriptsize cl}}_{\{P_{\mu}\}}$ 
such that $U\rho U^{\dagger}\not\in 
\mathcal{S}^{\mbox{\scriptsize
    cl}}_{\{P_{\mu}\}}$.
Since the partition $\{P_{\mu}\}$ is fine-grained,
it consists of one-dimensional 
projectors, $P_{\mu}=|\mu\rangle\langle\mu|$,  
where the vectors $|\mu\rangle$ form an orthonormal basis 
of  $\cal H$. The state $\rho$ can be written as $\rho=\sum_{\mu}p_ {\mu}
|\mu\rangle\langle\mu|$, where $p_{\mu}\ge0$ and 
$\sum_{\mu}p_ {\mu}=1$. The assumption $U\rho U^{\dagger}\not\in 
\mathcal{S}^{\mbox{\scriptsize
    cl}}_{\{P_{\mu}\}}$ implies that for at least one term in 
the decomposition $\,\rho=\sum_{\mu}p_ {\mu}
|\mu\rangle\langle\mu|\,$ classicality is not preserved. 
If it were not so, $U\rho U^{\dagger}$ would be classical.
Hence there exists $\mu_0$ such that $p_{\mu_0}\not=0$ and  
$(U |\mu_0\rangle\langle\mu_0|U^\dagger)
\not\in\mathcal{S}^{\mbox{\scriptsize cl}}_{\{P_{\mu}\}}$. 
This means there exist $\mu'$,$\mu''$, $\mu'\ne\mu''$, such that 
\begin{eqnarray}\label{superposition} 
\langle\mu'|U|\mu_0\rangle \equiv c_{\mu'} \ne 0 \;, \cr
\langle\mu''|U|\mu_0\rangle \equiv c_{\mu''} \ne 0 \;.
\end{eqnarray}

Now we derive a {\em necessary condition} for decoherence 
and then show that the above assumption contradicts it.
Written out, the decoherence
condition~(\ref{eq:decoherence}) is  
\begin{equation} \label{DecCondition}
\mbox{Tr}\left[P_{\alpha_k}UP_{\alpha_{k-1}}U\dots
  P_{\alpha_1}U\,\rho_0\: U^{\dagger}P_{\beta_1}\dots
  P_{\beta_{k-1}}U^{\dagger}P_{\beta_k}\right]\propto \prod_{j=1}^k
  \delta_{\alpha_j \beta_j}
\end{equation}
for all $k\in\natuerl$, all initial states 
$\rho_0\in\mathcal{S}^{\mbox{\scriptsize cl}}_{\{P_{\mu}\}}$,
and arbitrary histories $h_{\balphas}$,
$h_{\bbetas}$. By summing over $\alpha_2,\dots,\alpha_{k-1}$ and 
$\beta_2,\dots,\beta_{k-1}$, and using $\sum_{\mu}P_{\mu}=\one_{\cal H}$, 
we obtain
\begin{equation}\label{nec_con_(k-1)}
\mbox{Tr}\left[P_{\alpha_k}U^{k-1} P_{\alpha_1}U\,\rho_0\: U^{\dagger}
P_{\beta_1}(U^{\dagger})^{k-1}P_{\beta_k}\right]\propto
\delta_{\alpha_k\beta_k}
\delta_{\alpha_1\beta_1}
\end{equation}
for all $k\in\natuerl$, {any 
$\rho_0\in\mathcal{S}^{\mbox{\scriptsize cl}}_{\{P_{\mu}\}}$,} and
arbitrary $\alpha_1$, $\beta_1$, $\alpha_k$, $\beta_k$.

To derive a contradiction we let our histories start with 
the initial state $\rho_0=P_{\mu_0}\equiv|\mu_0\rangle\langle\mu_0|$.  
Furthermore we choose $\alpha_1=\mu'$,  
$\beta_1=\mu''$, and $\alpha_k=\beta_k=\mu_0$.
Since $\mu'\ne\mu''$, condition~(\ref{nec_con_(k-1)}) becomes
\begin{equation} \label{beforelemma}
%\forall k\in\natuerl\,:\; 
\mbox{Tr}\left[P_{\mu_0}U^{k-1} P_{\mu'}U\,\rho_0\: U^{\dagger}
P_{\mu''}(U^{\dagger})^{k-1}P_{\mu_0}\right] = 0 
\end{equation}
for all $k\in\natuerl$. 
On the other hand, since $\rho_0=|\mu_0\rangle\langle\mu_0|$, 
and using Eqs.~(\ref{superposition}),
we get for the left hand
side of Eq.~(\ref{beforelemma}):
\begin{equation} \label{beforelemma2}
\mbox{Tr}\left[P_{\mu_0}U^{k-1} P_{\mu'}U\,\rho_0\: U^{\dagger}
P_{\mu''}(U^{\dagger})^{k-1}P_{\mu_0}\right]
=\underbrace{c_{\mu'}c_{\mu''}^{*}}_{\not=0}\:\langle\,\mu''\,|
(U^{\dagger})^{k-1}P_{\mu_0}U^{k-1}|\,\mu'\,\rangle\;.
\end{equation}

We now make use of the Lemma. 
According to the Lemma, for any given, arbitrarily small $\epsilon>0$ 
we can always find a $q\in\natuerl$ such that 
$U^q=\one_{\cal H}+\hat{\mathcal{O}}(\epsilon)$, where 
$\hat{\mathcal{O}}(\epsilon)$ is some operator with norm bounded by 
$\epsilon$: 
$\parallel\!\!\hat{\mathcal{O}}(\epsilon)\!\!\parallel<\epsilon$.
Using the submultiplicativity property 
of operator norms, we have 
\begin{equation}
\parallel U^{-1} \hat{\mathcal{O}}(\epsilon)\parallel\,\le 
\,\parallel U^{-1}\parallel\times
\parallel \hat{\mathcal{O}}(\epsilon)\parallel\,=\,
\parallel \hat{\mathcal{O}}(\epsilon)\parallel
\end{equation}
and hence $U^{q-1}=U^{-1}+\hat{\mathcal{O}'}(\epsilon)$, where
$\parallel\! \hat{\mathcal{O}'}(\epsilon)\!\parallel<\epsilon$.
Choosing $k=q$ in Eq.~(\ref{beforelemma2}),
\begin{equation*}\hspace{-5cm}
\mbox{Tr}\left[P_{\mu_0}U^{q-1} P_{\mu'}U\,\rho_0\: U^{\dagger}
P_{\mu''}(U^{\dagger})^{q-1}P_{\mu_0}\right]
\end{equation*}
\begin{eqnarray} 
&=& c_{\mu'}c_{\mu''}^{*}\:\langle\,\mu''\,|
(U^{\dagger})^{q-1}P_{\mu_0}U^{q-1}|\,\mu'\,\rangle\nonumber\\
&=&
c_{\mu'}c_{\mu''}^{*}\:\langle\,\mu''\,|
\big(U+\hat{\mathcal{O}'}^{\dagger}(\epsilon)\big)\,|\mu_0\rangle\langle\mu_0|\,
\big(U^{\dagger}+\hat{\mathcal{O}'}(\epsilon)\big)\,|\,\mu'\,\rangle\nonumber\\
&=&
\underbrace{c_{\mu'}c_{\mu''}^{*}}_{\not=0}\:
\underbrace{\langle\,\mu''\,|
U\,|\mu_0\rangle}_{=c_{\mu''}}\underbrace{\langle\mu_0|\,
U^{\dagger}\,|\,\mu'\,\rangle}_{=c_{\mu'}^{*}} 
\:+ O(\epsilon)\nonumber\\
&=& \underbrace{|c_{\mu'}c_{\mu''}|^2}_{\not=0}
+ O(\epsilon)\;,
\end{eqnarray}
where $O(\epsilon)\to0$ as $\epsilon\to 0$.
This contradicts condition~(\ref{beforelemma}), 
which is a necessary consequence of our decoherence 
condition~(\ref{eq:decoherence}), 
and thus proves the theorem.

\noindent{\bf Proof of Theorem 2:}\newline
We prove theorem 2 by constructing a coarse-grained 
partition and a unitary map with the
required properties.  Let ${\cal H}$ be a 4-dimensional Hilbert space. We can
write ${\cal H}={\cal H}_{\mathcal{S}}\otimes{\cal H}_{\mathcal{E}}$, and
think of it as the Hilbert space of two qubits, regarding one of them 
as the system $\mathcal{S}$, the other one as the environment $\mathcal{E}$. 
Let $\{|0\rangle,|1\rangle\}$
and $\{|e_0\rangle,|e_1\rangle\}$ be orthonormal bases of ${\cal
  H}_{\mathcal{S}}$ and ${\cal H}_{\mathcal{E}}$, respectively. The states
$|\mu,e_{\lambda}\rangle:=|\mu\rangle\otimes |e_{\lambda}\rangle$, where
$\mu,\lambda\in\{0,1\}$, form an orthonormal basis of ${\cal H}$.

We now define a coarse-grained projective partition, 
$\{P_0,P_1\}$, by
\begin{eqnarray} 
P_{\mu}&=&|\mu\rangle\langle\mu|\otimes\one_{\cal H_{\mathcal{E}}}\nonumber\\
&=&|\mu,e_0\rangle\langle\mu, e_0|+
|\mu,e_1\rangle\langle\mu, e_1|\;,
\end{eqnarray} 
and a unitary map 
$U:\cal H\rightarrow\cal H$ by
\begin{eqnarray} 
U|0\,,e_0\rangle&=&|0\,,e_1\rangle\cr
U|0\,,e_1\rangle&=&|1\,,e_0\rangle\cr
U|1\,,e_0\rangle&=&|1\,,e_1\rangle\cr
U|1\,,e_1\rangle&=&|0\,,e_0\rangle\;.
\end{eqnarray}
The map $U$ is a permutation of the basis states. 
A more compact definition of $U$ is  
\begin{equation}
U|\mu,e_{\lambda}\rangle=\sum_{\nu=0}^1 \delta_{\nu\lambda} 
|\mu+\nu,e_{1+\nu}\rangle \;,
\end{equation}
where $\mu,\lambda\in\{0,1\}$ and addition is understood modulo 2. The map $U$
does not preserve classicality w.r.t.\ $\{P_0,P_1\}$. This can be seen by
considering the pure classical state $\rho=\Ket\psi\langle\psi|
\in\mathcal{S}^{\mbox{\scriptsize cl}}_{\{P_{\mu}\}}$, where
$|\psi\rangle=\frac{1}{\sqrt{2}}(|0,e_0\rangle+
|0,e_1\rangle)\in\mbox{supp}(P_0)$. Since 
$U|\psi\rangle=\frac{1}{\sqrt{2}} (|0,e_1\rangle+|1,e_0\rangle)$ is a
superposition of states that belong to ${\rm supp}(P_0)$ and ${\rm
  supp}(P_1)$, respectively, we have
$U\rho U^{\dagger}\not\in \mathcal{S}^{\mbox{\scriptsize cl}}_{\{P_{\mu}\}}$.
This  shows that, with our choice of unitary map and partition, 
the classicality condition~(\ref{eq:classicality}) is not satisfied.

It remains to be shown that the decoherence condition~(\ref{eq:decoherence})
is satisfied for this choice of unitary map and partition.  The most general
classical state w.r.t.\ $\{P_0,P_1\}$ is given by $\rho= p_0\rho_0+p_1\rho_1$,
where $p_0+p_1=1$ and $\rho_0$, $\rho_1$ are any density matrices satisfying
$\mbox{supp}(\rho_\mu) \subseteq\mbox{supp}(P_\mu)$ for $\mu=0,1$.
Let $\rho_{\mu}=\sum_{j=0}^1
r_{\mu}^j |\omega_{\mu}^j\rangle\langle\omega_{\mu}^j|\,$ be their spectral
decompositions. In terms of the basis vectors
$|\mu,e_{\lambda}\rangle$ the eigenvectors can be 
written as $\, |\omega_{\mu}^j\rangle=\sum_{\lambda =0}^1
c_{\mu,\lambda}^j|\mu\,,e_{\lambda}\rangle$. Putting everything together, we 
find that every $\rho\in\mathcal{S}^{\mbox{\scriptsize cl}}_{\{P_{\mu}\}}$
can be written in the form
\begin{equation}
\rho=\sum_{\mu=0}^1\sum_{j=0}^1\sum_{\lambda,\lambda'=0}^1
\,p_{\mu}r_{\mu}^jc_{\mu,\lambda}^jc_{\mu,\lambda'}^{j\,*}
|\mu\,,e_{\lambda}\rangle\langle\mu\,,e_{\lambda'}|  \;.
\end{equation}
Substituting this into the expression 
\begin{equation}
\mathcal{R}_{\rho}(k):=P_{\alpha_k}UP_{\alpha_{k-1}}U\dots
  P_{\alpha_1}U\,\rho\: U^{\dagger}P_{\beta_1}\dots
  P_{\beta_{k-1}}U^{\dagger}P_{\beta_k}
\end{equation}
and using the principle of induction, one can show that, for $k\ge2$ and
any $\rho\in\mathcal{S}^{\mbox{\scriptsize cl}}_{\{P_{\mu}\}}$,
\begin{eqnarray} 
\mathcal{R}_{\rho}(k) \propto
\delta_{\alpha_1+\alpha_{k-1}+\alpha_{k},
\beta_1+\beta_{k-1}+\beta_{k}}
|\alpha_k,e_{\alpha_{k-1}+\alpha_{k}+1}\rangle\langle
\beta_k,e_{\beta_{k-1}+\beta_{k}+1}|\;,
\end{eqnarray}
where again addition is understood modulo 2.
This can be shown to be equivalent to 
\begin{equation}
\mathcal{R}_{\rho}(k)\propto\left(\prod_{j=1}^{\lfloor\frac{k-1}{2}\rfloor}
\delta_{\alpha_1+\alpha_{2j+1},\beta_1+\beta_{2j+1}}\right)
\left(\prod_{j=1}^{\lfloor\frac{k}{2}\rfloor}
\delta_{\alpha_{2j},\beta_{2j}}\right)
|\alpha_k,e_{\alpha_{k-1}+\alpha_{k}+1}\rangle\langle
\beta_k,e_{\beta_{k-1}+\beta_{k}+1}|\;.
\end{equation}
Taking the trace on both sides gives
\begin{equation}
\mbox{Tr}[\mathcal{R}_{\rho}(k)]
\propto\left(\prod_{j=1}^{\lfloor\frac{k-1}{2}\rfloor}
\delta_{\alpha_1+\alpha_{2j+1},\beta_1+\beta_{2j+1}}\right)
\left( \prod_{j=1}^{\lfloor\frac{k}{2}\rfloor}
\delta_{\alpha_{2j},\beta_{2j}}\right)
\langle\beta_k,e_{\beta_{k-1}+\beta_{k}+1}|
\alpha_k,e_{\alpha_{k-1}+\alpha_{k}+1}\rangle\;.
\end{equation}
The scalar product on the right hand side is equal to 
$\delta_{\alpha_k,\beta_k}\delta_{\alpha_{k-1},\beta_{k-1}}$. 
Using
\begin{equation}
\delta_{\alpha_k,\beta_k}\delta_{\alpha_{k-1},\beta_{k-1}}
\prod_{j=1}^{\lfloor\frac{k-1}{2}\rfloor}
\delta_{\alpha_1+\alpha_{2j+1},\beta_1+\beta_{2j+1}}
=\prod_{j=1}^{\lfloor\frac{k-1}{2}\rfloor}
\delta_{\alpha_{2j+1},\beta_{2j+1}}  \;,
\end{equation}
and the fact that 
$\mbox{Tr}[\mathcal{R}_{\rho}(k=1)]\propto\delta_{\alpha_{1},\beta_{1}}$, 
we finally obtain
\begin{equation}
\mbox{Tr}[\mathcal{R}_{\rho}(k)]
\propto\prod_{j=1}^k\delta_{\alpha_{j},\beta_{j}}
\end{equation}
for all $k\in\natuerl$ and all
$\rho\in\mathcal{S}^{\mbox{\scriptsize cl}}_{\{P_{\mu}\}}$. The decoherence
condition~(\ref{eq:decoherence}) is thus satisfied, which completes the proof
of Theorem 2.

\vspace{2mm}

Let us conclude with a brief discussion.  We have derived a simple, necessary
and sufficient, decoherence condition for sets of fine-grained histories of
arbitrary length. To verify that our condition holds for a particular unitary
map $U$, only a single iteration of the map has to be taken into account,
which can be much easier than establishing decoherence directly by computing
the off-diagonal elements of the decoherence functional.  This is
especially useful for studying chaotic quantum maps, for which typically only
the first iteration is known in  closed analytical form
\cite{Soklakov2000a}. 

Our results can be summarized as follows. We have analyzed the relationship
between the condition that a unitary map is classicality-preserving on the one
hand, and the decoherence condition for all classical initial states and
arbitrarily long histories on the other hand. We have shown that for
fine-grained histories, these two conditions are equivalent, but that
decoherence of coarse-grained histories does not, in general, imply 
that the unitary evolution preserves classicality of states.

\noindent{\bf Acknowledgment:} We would like to thank Todd Brun for helpful
discussions.

\end{document}